\documentclass{webofc}

\usepackage[varg]{txfonts}   
\usepackage{hyperref}
\usepackage{url}
\hypersetup{colorlinks=true,citecolor=blue,urlcolor=blue,linkcolor=blue}
\usepackage{amsmath, amssymb}
\usepackage{bm}
\usepackage{caption}
\usepackage{subcaption}
\usepackage{float}
\usepackage{booktabs}

\usepackage{graphicx}
\usepackage{caption}
\usepackage{array}

\newcommand{\PrayInit}{P_{\mathrm{ray, init}}}
\newcommand{\PrayInc}{P_{\mathrm{ray, inc}}}
\newcommand{\Pbeam}{P_\mathrm{tot}}
\newcommand{\Nrays}{n_\mathrm{rays}}
\newcommand{\Nseg}{n_\mathrm{seg}}
\newcommand{\rayposfront}{\bm r_\mathrm{front}}
\newcommand{\rayposback}{\bm r_\mathrm{back}}
\newcommand{\rayposvol}{\bm r_\mathrm{vol}}
\newcommand{\raydir}{\bm d}
\newcommand{\lseg}{l_\mathrm{seg}}

\newcommand{\absorbfacV}{\alpha_\mathrm{V}}
\newcommand{\absorbfacS}{\alpha_\mathrm{S}}
\newcommand{\Iinit}{I_\mathrm{init}}
\newcommand{\I}{I}
\newcommand{\Pray}{P_{\mathrm{ray}}}

\newcommand{\heatfrontsurf}{q_\mathrm{frontsurf}}
\newcommand{\heatbacksurf}{q_\mathrm{backsurf}}
\newcommand{\heatvol}{q_\mathrm{vol}}

\newcommand{\nel}{n_{el}}
\newcommand{\nnodestot}{n_\mathrm{tn}}
\newcommand{\nnodesel}{n_\mathrm{en}}
\newcommand{\Le}{\bm L^{e}}
\newcommand{\tempel}{T^e}
\newcommand{\tempelvec}{\bm T^e}
\newcommand{\xvec}{\bm x}
\newcommand{\xivec}{\bm \xi}
\newcommand{\xiposvol}{\bm \xivec_\mathrm{vol}}
\newcommand{\afuncvec}{\bm N}
\newcommand{\afunc}{N}
\newcommand{\xel}{\bm x^e}
\newcommand{\yel}{\bm y^e}
\newcommand{\zel}{\bm z^e}

\newcommand{\Kt}{\bm K_{t}}
\newcommand{\Dt}{\bm D_{t}}
\newcommand{\flux}{\bm f}
\newcommand{\tempvec}{\bm T}

\newcommand{\heatloadvecel}{\bm f^e}
\newcommand{\heatloadsingle}{f}
\newcommand{\elphysvol}{\Omega^e}
\newcommand{\elphysinfvol}{\Omega}
\newcommand{\elrefvol}{\tilde{\Omega}^e}
\newcommand{\elrefinfvol}{\tilde{\Omega}}

\newcommand{\heatsourcevol}{s}

\newcommand{\jacobiandet}{J}
\newcommand{\jacobianmatrix}{\bm J}
\newcommand{\function}{\bm f}
\newcommand{\deltaxi}{\Delta \xivec}
\newcommand{\error}{\epsilon}

\newcommand{\distance}{d}
\newcommand{\ndist}{n_\mathrm{dist}}

\newcommand{\wavelength}{\lambda}

\begin{document}
\title{Versatile Absorption Modeling for Transmissive Optical Elements Using Ray Tracing and Finite Element Analysis}
%
%

\author{\firstname{Mark} \lastname{Kurcsics}\inst{1}\fnsep\thanks{\email{mark.kurcsics@itm.uni-stuttgart.de}} \and
        \firstname{Peter} \lastname{Eberhard}\inst{1}\fnsep\thanks{\email{peter.eberhard@itm.uni-stuttgart.de}} 
}

\institute{Institute of Engineering and Computational Mechanics, University of Stuttgart, Pfaffenwaldring 9, 70569 Stuttgart, Germany }

\abstract{Emerging quality requirements in modern optical systems increase the need for accurate simulation and compensation of disturbances. One key disturbance in high-power laser applications results from thermal loading due to absorption. This paper presents a flexible modeling approach for absorption based on ray tracing, where point heat sources are defined along ray paths. In addition, several methods for mapping these point heat sources along the ray path to the finite element mesh are introduced and compared. A major advantage of the proposed approach lies in its flexibility, as it enables the simulation of transient phenomena such as dynamic beam positioning and time-dependent variations in beam shape.}
\maketitle
\section{Introduction and Motivation}
\label{intro}

Modern optical systems, such as those used in high-power laser applications, are highly complex and demand exceptional process stability. Various factors can impair their performance, with thermal disturbances being particularly significant. These disturbances may arise from ambient temperature changes or, in the case of high-power lasers, from partial absorption of the laser beam within optical elements, which leads to temperature gradients inside the material \cite{GorajekEtAl20}. These gradients result in changes to the refractive index and induce thermal deformations, both of which degrade optical performance.

Although analytical models exist to describe these effects \cite{Bisson09, BaessoShenSnook94, GlurLaviGraf04, Koechner70}, they rely on strong simplifications, limiting their applicability in complex, real-world systems \cite{Rall23, GorajekEtAl20}. At the same time, numerical simulation has become an essential and effective tool. Simulations enable design evaluations early in the development process, helping to reduce costs by minimizing the need for late-stage adjustments.

Ray tracing has become the standard technique for simulating optical behavior, while the finite element method is widely used for thermal analysis. Coupling these methods provides a robust foundation for modeling absorption effects in optical elements. Ideally, a numerical algorithm for computing absorption should be flexible enough to account for arbitrary beam shapes, distortions induced by optical components, and time-dependent variations in beam shape and position.

Several approaches have been proposed in the literature to address this challenge. In \cite{Bonhoff19} an approach is presented based on computing the heat flux on an auxiliary mesh, which is then mapped to the finite element mesh. The method supports both rotationally symmetric and non-symmetric static beam profiles. However, it is not suitable for modeling time-dependent variations. Another method is proposed by \cite{YuEtAl16} that accounts for intensity distributions by using ray paths to define individual heat sources, which are mapped to the finite element mesh. However, heat distribution within a lens is calculated only through linear interpolation, and time-dependent effects are not considered. The algorithm described by  \cite{QinEtAl12} is developed for laser welding. It applies ray tracing with each ray's power being proportional to the local beam intensity. However, since the method is designed for modeling absorption on metal surfaces, it is not applicable to optical elements. The method presented by \cite{LyuZhan21} is a coupled ray tracing-finite element approach. Heat sources are calculated and transferred to the finite element mesh, but the contribution to each element is averaged. However, this averaging requires a fine thermal mesh to be able to resolve gradients in the heat flux. A similar strategy is presented by \cite{GenbergMichelsBisson19}, where heat sources are computed along ray paths. An additional voxel mesh is introduced, and heat deposition per voxel is mapped to the finite element mesh. This extra additional mesh introduces additional the complexity to the absorption method. The method introduced by \cite{CossonEtAl19} computes absorbed energy within volume elements. The power of each ray is determined based on its initial position relative to a Gaussian beam profile. However, their focus is on inhomogeneous materials which necessitates to introduces macroscopic and microscopic meshes. An absorption model for selective laser melting is implemented by \cite{DorussenGeersRemmers23}. Rays are spatially distributed according to a normal distribution, and each ray receives a fraction of the total laser energy. As the algorithm is tailored to absorption in metallic particles, it is not applicable to optical elements. The method proposed by \cite{RallEtAl22} incorporates an auxiliary mesh for intensity distribution. However, the mesh is reconstructed after tracing rays through the optical system, with absorption effects taken into account. The approach presented by \cite{HahnEberhard22} prescribes a heat flux for each ray on the surface intersection point and performs linear interpolation for obtaining the heat flux within the volume. However, no physical law is involved for calculating the heat flux.

In this paper, we present a novel approach for modeling light absorption in transmissive optical elements that directly couples ray tracing with the finite element method. Our method offers high flexibility in handling complex and time-dependent beam profiles enabling simulations of thermally induced effects, such as thermal deformation and spatially varying refractive index, in modern optical systems.

In Sec.~\ref{sec:methods} the methods are presented, implying the ray tracing method (Sec.~\ref{sec:raytracing}), the finite element method (Sec.~\ref{sec:fem}), the absorption model (Sec.~\ref{sec:absorption}) and the methods for mapping the point heat sources (Sec.~\ref{sec:mapping}). In Sec.~\ref{sec:results} the simulation setup and the results are presented. The paper closes with Sec.~\ref{sec:conclusion} where the conclusions are presented.

\section{Methods}
\label{sec:methods}

\subsection{Simulative procedure}

The overarching simulation approach involves a sequential, multi-step process designed to model the thermal response of the optical element to absorption effects. The first step is to perform a transient ray tracing of the thermally undisturbed system for the complete simulation duration, as described in Sec.~\ref{sec:raytracing}, which simulates the propagation of rays through the optical system and determines their interactions with surfaces and materials. The spatial distribution of ray positions and ray paths obtained from this step serve as the basis for calculating the heat sources resulting from energy absorption, detailed in Sec.~\ref{sec:absorption}. These heat sources are computed by evaluating the energy deposited as the rays interact with surface and bulk material, taking into account heat absorption.

Once the absorption-induced heat sources have been identified and the heat originating form each of these is determined, the next step is to map these discrete heat values onto the nodes of the finite element mesh (see Sec.~\ref{sec:mapping}). This mapping process is crucial for ensuring that the energy input is represented within the finite element framework. With the heat sources distributed across the mesh, the final stage of the simulation involves solving for the resulting temperature field for each timestep of the simulation. This is accomplished using the thermal finite element method (Sec.~\ref{sec:fem}), which computes the temperature distribution based on the imposed heat flux and the governing equations of heat transfer. Together, these steps form an integrated simulation pipeline that enables dynamic thermal analysis based on ray-based energy deposition.

\subsection{Ray tracing}
\label{sec:raytracing}

A commonly used technique for optical analysis is ray tracing, which models the propagation of light through an optical system by discretizing the light source into a large number of individual rays. These rays are then traced through the system using principles of vector algebra and geometrical optics \cite{Gross05}, allowing for a detailed simulation of light behavior such as reflection, refraction, and absorption.

The mathematical basis for ray propagation in general form is described by a differential equation
\begin{equation}
\frac{\mathrm{d}}{\mathrm{d} s} \left( n(\bm r) \frac{\mathrm{d} \bm r}{\mathrm{d} s} \right) = \frac{\partial n}{\partial s} \quad .
\end{equation}
where $n(\bm r)$ denotes the refractive index at a given position $\bm r$, and $s$ is the geometric path length, which represents the physical distance traveled by the ray \cite{Stoerkle18}. This formulation accounts for the effects of spatial inhomogeneity in the refractive index, which becomes particularly important in gradient-index (GRIN) media. In such cases, the light path is curved due to variations in refractive index, for example, as a result of temperature gradients or material composition changes.

In contrast, for homogeneous media where the refractive index is constant throughout the domain, the equation simplifies significantly, and rays propagate in straight lines.  

There are various implementations of ray tracing. In sequential ray tracing, for instance, the interaction of each ray with the optical surfaces is computed in the specific order in which the elements are arranged in the system. At each interface, the ray is typically refracted or reflected based on the vector form of Snell’s law, which determines the change in direction according to the refractive indices of the two media and the angle of incidence.

As discussed in Sec.~\ref{sec:absorption}, the ray tracing algorithm must also account for the spatial intensity distribution of the light source. This is particularly important when modeling physically realistic sources. For example, if the light source emits a Gaussian beam, the rays should be distributed in space according to a Gaussian probability density function. This ensures that the ray density, and thus the simulated light intensity, faithfully represents the actual beam profile.

\subsection{Thermal Finite Element Method}
\label{sec:fem}

The finite element method is a widely adopted numerical technique used for solving partial differential equations across various physical domains, including structural mechanics, fluid dynamics, and heat transfer. In the context of this study, the focus lies on thermal phenomena, and thus the thermal finite element method is employed to compute temperature fields resulting from localized heat sources.

At its core, the finite element method involves discretizing a physical body into a finite number of elements, which are connected at nodes. These elements can take various shapes, such as triangles, quadrilaterals, or tetrahedra, depending on the geometry and dimensionality of the domain. The nodes, which form the vertices of the elements, play a central role in the numerical solution. The primary unknown, the temperature in this case, is computed at these nodes, and the field within each element is approximated using interpolation functions (shape functions) based on the nodal values.

The governing heat conduction equation, typically a second-order partial differential equation in space and first-order in time, is transformed into a system of ordinary differential equations through spatial discretization into $\nel$ elements and a total of $\nnodestot$ nodes
\begin{equation}
\Dt \dot{\tempvec}(t) + \Kt \tempvec(t) = \flux(t)
\end{equation}
with the capacitance matrix $\Dt \in \mathbb{R}^{\nnodestot \times \nnodestot}$, the conductance matrix $\Kt \in \mathbb{R}^{\nnodestot \times \nnodestot}$, the time-dependent flux vector $\flux(t) \in \mathbb{R}^{\nnodestot \times 1}$, and the time-dependent temperature vector $\tempvec(t) \in \mathbb{R}^{\nnodestot \times 1}$ \cite{Nicholson08, FishBelytschko07}. This differential equation is then solved for the temperature vector $\tempvec(t)$, which provides the temporal evolution of nodal temperatures.

In detail, the flux vector $\flux(t)$ is the excitation term of the differential equation. It includes the effects of volumetric heat sources and surface heat sources and defines the heat input for each node. Here, the flux is governed by the volumetric absorption as well as the absorption due to anti-reflection coatings on the surface. The global heat flux vector $\flux$ is composed through summation over the element heat flux vectors $\heatloadvecel$. Thereby, the element heat flux $\heatloadvecel$ is mapped to the corresponding globally numbered nodes through the element gather matrices $\Le \in \mathbb{R}^{\nnodestot \times \nnodesel}$ \cite{FishBelytschko07}
\begin{equation}
\flux = \sum_{e = 1}^{\nel} \Le \heatloadvecel \quad .
\end{equation}

While gather matrices provide a good conceptual basis for understanding the summation process and the assembly of global vectors and matrices, it is numerically more efficient in implementation to use lookup tables, where for each local node in an element the corresponding global node number is stored. These lookup tables are typically used in computational frameworks to speed up the matrix assembly process and minimize overhead, especially in large-scale simulations.

Overall, the thermal finite element method offers a robust and flexible framework for modeling transient and steady-state heat transfer problems. Its ability to handle complex geometries and heterogeneous material properties makes it particularly suitable for simulations involving spatially distributed heat sources, such as those arising from optical absorption in this study.

\subsection{Absorption Modeling}
\label{sec:absorption}

In optical systems incorporating lenses with anti-reflective coatings, two primary absorption mechanisms contribute to heat generation. The first is surface absorption due to the absorptive properties of the coating, while the second is volumetric absorption within the lens material itself \cite{Gatej14}. The latter is typically modeled using Beer–Lambert’s law \cite{PiehlerEtAl12, HuegelGraf09}, which describes the exponential attenuation of an initial light intensity $\Iinit$ as it traverses a medium of thickness $z$ with a material-specific absorption coefficient $\absorbfacV$
\begin{equation}
\I(x, y, z) = \Iinit(x, y) e^{- \absorbfacV z} \; .
\end{equation}

The goal of the proposed absorption algorithm is to bridge ray tracing and thermo-mechanical finite element method in a flexible and generalizable manner. Since the Beer-Lambert's law is inherently tied to the local intensity of the light field, the strategy involves distributing rays spatially to represent the beam's intensity profile. For instance, in the case of a Gaussian beam, this means sampling ray positions such that the ensemble reflects the Gaussian distribution of intensity.

To each individual ray an initial power $\PrayInit$ is assigned such that the total beam power $\Pbeam$ is distributed across $\Nrays$ rays
\begin{equation}
\PrayInit = \frac{\Pbeam}{\Nrays} \;.
\end{equation}

By preserving the intensity distribution through spatial ray placement and equal power assignment, the algorithm remains adaptable to arbitrary initial beam shapes and accounts for beam reshaping by refractive elements.

After initialization, each ray is traced through the optical system to determine its propagation path and interaction points (see Sec.~\ref{sec:absorption}). Subsequently, the absorbed heat along each ray is computed individually.

In the following, the absorption process is described in the context of a single lens. When a ray strikes the front surface of the lens, a fraction of its power is absorbed by the anti-reflective coating. This surface absorption is modeled using a scalar surface absorption factor $\absorbfacS$, which serves as a lumped parameter representing complex physical dependencies that are difficult to model explicitly \cite{Bonhoff19, PiehlerEtAl12}. The resulting point heat source on the surface $\heatfrontsurf$ is computed as
\begin{equation}
\heatfrontsurf = \absorbfacS \PrayInc \; .
\end{equation}

The incoming ray power $\PrayInc$ equals the initial ray power $\PrayInit$ for the first lens; for subsequent optical elements, $\PrayInc$ is updated to account for prior absorptive losses. The surface heat source $\heatfrontsurf$ is then mapped to the surrounding finite element nodes using the method described in Sec.~\ref{sec:mapping}.

After traversing the front surface, the remaining ray power undergoes volumetric attenuation according to Beer-Lambert's law. The residual power after passing through a lens is
\begin{equation}
\Pray(l) = (\PrayInc - \heatfrontsurf) e^{- \absorbfacV l} \; , 
\end{equation}
where $l$ is the geometric path length within the lens. Because the absorption occurs continuously along the ray path, it must be discretized for compatibility with the finite element model. 
To this end, the ray path between its front entry point $\rayposfront$ and back exit point $\rayposback$ is subdivided into $\Nseg$ segments of equal length $\lseg$
\begin{equation}
\lseg = \frac{\lVert \rayposback - \rayposfront \lVert_2}{\Nseg} \; .
\end{equation}

The absorbed heat in each segment $i \in [1, \Nseg]$ is computed recursively. Let ${\Pray}_i$ denote the ray power entering segment $i$, and ${\heatvol}_i$ the volumetric heat absorbed in that segment. The computational scheme is then
\begin{equation}
\begin{aligned}
    {\Pray}_i &= {\Pray}_{i-1} - {\heatvol}_{i-1} \;, \\
    {\heatvol}_i &= {\Pray}_i \left(1 - \mathrm{e}^{- \absorbfacV \lseg} \right) \;, \\
    {\Pray}_0 &= \PrayInc - \heatfrontsurf \;, \quad {\heatvol}_0 = 0 \; .
\end{aligned}
\end{equation}

Each volumetric heat flux contribution is modeled as a point heat source located at the segment center
\begin{equation}
{\rayposvol}_i = \rayposfront + \frac{2i-1}{2} \lseg \raydir \quad \text{with} \quad \raydir = \frac{\rayposback - \rayposfront}{\lVert \rayposback - \rayposfront \lVert_2} \;.
\end{equation}

These discrete point heat sources are then mapped to the finite element mesh using appropriate spatial weighting schemes as presented in Sec.~\ref{sec:mapping}.

At the back surface due to the coating, energy is absorbed again. The resulting point heat source is given by
\begin{equation}
     \heatbacksurf = \absorbfacS \left(  {\Pray}_{\Nseg} - {\heatvol}_{\Nseg}  \right)  \; .
\end{equation}

\subsection{Heat Source Mapping}
\label{sec:mapping}

\subsubsection{Mapping based on Shape functions}

The mapping of heat sources into the finite element context is essential for transferring ray-tracing-based energy deposition into the thermal simulation domain. This section describes how discrete point heat sources, derived from optical ray tracing and optical absorption, are projected onto the finite element mesh using an isoparametric formulation.

In the isoparametric approach, both the geometry and the field variables within an element $e$ with $\nnodesel$ nodes are interpolated using the same shape functions $\afuncvec(\xivec) = [\afunc_1(\xivec) \; \dots \; \afunc_{\nnodesel}(\xivec)] \in \mathbb{R}^{1 \times \nnodesel}$, which are defined in terms of local coordinates $\xivec = [\xi \; \eta \; \zeta]^\top \in \mathbb{R}^{3 \times 1}$ with $-1 \leq \xi,\eta,\zeta \leq 1$ \cite{FishBelytschko07}. The temperature field $\tempel$ and spatial coordinates $\xvec$ are thus represented as
\begin{align}
\tempel(\xivec) &= \afuncvec(\xivec) \tempelvec \in \mathbb{R} \; , \\
\xvec(\xivec) &=
\begin{bmatrix}
\afuncvec(\xivec) \xel \\
\afuncvec(\xivec) \yel \\
\afuncvec(\xivec) \zel
\end{bmatrix}
\in \mathbb{R}^{3 \times 1} \; ,
\end{align}
where $\xel, \yel, \zel \in \mathbb{R}^{\nnodesel \times 1}$ are the global coordinates of the element's nodes in the $x$, $y$, and $z$ directions and $\tempelvec \in \mathbb{R}^{\nnodesel \times 1}$ is the vector of the nodal temperatures.

Given a point heat source ${\heatvol}$ located at position ${\rayposvol}$, its contribution to the nodal heat flux must be distributed across the finite element mesh. Since the finite element formulation is inherently element-wise, it is first necessary to determine which element contains the point source. A simple strategy is to locate the closest node to ${\rayposvol}$ and check the adjacent elements. For each candidate element, the global coordinates ${\rayposvol}$ are mapped to the local coordinates $\xiposvol = \xivec(\rayposvol)$, and containment is confirmed if each component of $\xiposvol$ lies within the standard domain $[-1, 1]$.

The forward mapping from local to global coordinates, $\xvec(\xivec)$, is well-defined, straightforward to evaluate numerically but in most cases nonlinear. Therefore, the inverse mapping $\xivec(\xvec)$ is typically not analytically available \cite{SchraderEtAl14}. To compute $\xiposvol$, a Newton–Raphson iteration is employed by solving the nonlinear residual equation
\begin{equation}
\function(\xivec) = \xvec(\xivec) - \rayposvol 
\end{equation}
using the iterative update
\begin{equation}
\jacobianmatrix(\xivec_n) \deltaxi_n = - \function(\xivec_n) \; , \; \text{where} \;
\jacobianmatrix(\xivec) = \frac{\partial \function(\xivec)}{\partial \xivec} =  \frac{\partial \xvec(\xivec)}{\partial \xivec} \;, \;
\xivec_{n+1} = \xivec_n + \deltaxi_n \;, \; \xivec_0 = \bm 0  \; .
\end{equation}
The iteration continues until the residual norm satisfies the convergence criterion
\begin{equation}
\lVert \function(\xivec_{n+1}) \rVert_2 < \error \; .
\end{equation}

Once the containing element is identified, the heat source ${\heatvol}$ is modeled as a volumetric heat source density acting over an infinitesimally small volume, represented using the Dirac delta function centered at the source location
\begin{equation}
\heatsourcevol = {\heatvol} \delta(\xvec - {\rayposvol}) \; .
\end{equation}
The elemental heat flux vector $\heatloadvecel$ induced is given by
\begin{equation}
\heatloadvecel = \int_{\elphysvol} \afuncvec^\top \heatsourcevol \mathrm{d} \elphysinfvol
= \int_{\elphysvol} \afuncvec^\top {\heatvol}  \delta(\xvec - {\rayposvol})  \mathrm{d} \elphysinfvol .
\end{equation}
Applying the isoparametric transformation to switch from global to local coordinates results in
\begin{equation}
\heatloadvecel = \int_{\elrefvol} \afuncvec^\top(\xivec) {\heatvol}  \delta(\xvec(\xivec) - {\rayposvol})  |\jacobiandet(\xivec)| \mathrm{d} \elrefinfvol 
\end{equation}
where the Jacobian determinant $\jacobiandet(\xivec) = \det \left( \frac{\partial \xvec}{\partial \xivec} \right)$ accounts for the transformation of volume.
As outlined by \cite{SaichevWoyczynski18}, the delta function can be reformulated in terms of local coordinates
\begin{equation}
    \delta(\xvec(\xivec) - {\rayposvol}) = \frac{ \delta(\xivec - \xivec({\rayposvol}))}{|\jacobiandet(\xivec)| } = \frac{ \delta(\xivec -{\xiposvol})}{|\jacobiandet(\xivec)| } \;.
\end{equation}

The local coordinates  ${\xiposvol} = \xivec({\rayposvol})$ are reused from the determination of the element membership as described earlier. Then, the integral results in 
\begin{equation}
     \heatloadvecel = \int_{\elrefvol} \afuncvec^\top(\xivec) {\heatvol} \delta(\xivec -{\xiposvol}) \mathrm{d} \elrefinfvol \;.
\end{equation}
Then, the sifting property of the Dirac delta yields the heat flux vector
\begin{equation}
     \heatloadvecel = \afuncvec^\top(\xiposvol) {\heatvol} \;.
\end{equation}

This provides the nodal distribution of the point heat source ${\heatvol}$ within the containing element. The resulting local heat flux vector $\heatloadvecel$ is then assembled into the global heat flux vector $\flux$ based on the global node indices of the element nodes as explained in Sec.~\ref{sec:fem}.

\subsubsection{Mapping Based on Global Inverse Distances}

Since the procedure involving shape functions requires the use of the Newton-Raphson method to compute the local coordinates, it introduces iterative computations that can be computationally expensive, especially when handling a large number of point heat sources. To reduce the computational effort, an alternative approach is proposed that bypasses the need for local coordinate mapping. This method is based on a weighted inverse distance mapping of point heat sources directly to nearby nodes of the finite element mesh. It saves computation time at the expense of being less accurate.

In this approach, for each point heat source located at position $\rayposvol$, the Euclidean distance to the closest mesh node is first computed. If this distance is exactly zero, meaning the heat source coincides with a mesh node, the full heat contribution $\heatvol$ is assigned directly to that node, eliminating the need for further distribution. However, in the more general case where the heat source lies between mesh nodes, the heat must be distributed across several nearby nodes to reflect its spatial influence.

To achieve this, the Euclidean distances $\distance_j$ from the heat source to a selected number $\ndist$ of neighboring nodes with $j \in [1,2,\dots,\ndist]$ are computed. The heat is then distributed to these nodes inversely proportional to their distance from the source, such that closer nodes receive a larger share of the heat. The heat flux contribution to each node is given by
\begin{equation}
\heatloadsingle_j = \frac{\distance_j^{-1} \heatvol}{\sum\limits_{k=1}^{\ndist} \distance_k^{-1}} \; .
\end{equation}
This formulation ensures conservation of energy, as the total distributed heat sums to $\heatvol$. Additionally, the inverse-distance weighting yields a smooth and physically plausible distribution without requiring knowledge of the element shape functions or iterative solving.

Finally, the computed nodal contributions $\heatloadsingle_j$ are assembled into the global heat flux vector $\flux$ by adding them to the corresponding global node indices. This approach significantly simplifies the implementation and provides, therefore, a practical compromise between computational simplicity and spatial accuracy.

\subsubsection{Mapping Based on Element-wise Inverse Distances}

A third possibility is to combine both previously discussed approaches by mapping the point heat source to the nodes of an enclosing finite element using inverse-distance weighting. This hybrid method aims to retain the local resolution of shape-function-based mapping while avoiding the complexity of iterative schemes. For each point heat source, the closest node is identified first. If the heat source position exactly coincides with a node, the corresponding heat $\heatvol$ is directly assigned to that node.
Otherwise, the elements are determined that include the closest node. To identify the potential enclosing element, Delaunay triangulation is used to connect the nodes for each neighboring element. Each candidate element is then tested to check whether it contains the heat source position. Once the enclosing element is found, the heat is distributed to its $\nnodesel$ nodes using inverse-distance weights
\begin{equation}
\heatloadsingle_j = \frac{\distance_j^{-1} \heatvol}{\sum\limits_{k=1}^{\nnodesel} \distance_k^{-1}} \; , \; j \in [1,2, \dots, \nnodesel] \; ,
\end{equation}
where $\distance_j$ denotes the Euclidean distance from the heat source to node $j$. As in the previous strategies, the individual nodal heat loads $\heatloadsingle_j$ are assembled into the global flux vector $\flux$ based on the global node numbering.

\section{Simulation Setup and Results}
\label{sec:results}

In the following, the behavior of the absorption algorithm and associated heat mapping procedures is evaluated using a simplified scenario involving a single plano-convex N-BK7 lens. The focus of the simulations is on examining the qualitative behavior. However, realistic material parameters are chosen. Exemplarily, for N-BK7 at a wavelenght of $\wavelength =  1.064~\mu\mathrm{m}$ the volumetric absorptivity is $\absorbfacV = 0.1286~\frac{1}{\mathrm{m}}$ \cite{Polyanskiy24}. For the surface absorptivity $\absorbfacS = 50\cdot 10^{-6}$ \cite{Bonhoff19} is selected. For the finite element mesh isoparametric hexahedral elements are chosen. As initial temperature for the lens $20 ^{\circ} \mathrm{C}$ is preset as well as a fixed temperature boundary condition of $20 ^{\circ} \mathrm{C}$ at the outer cylindrical surface of the lens. Heat generation is computed for the beam path and mapped to the mesh using different strategies (using shape functions, global inverse-distances and element-wise inverse distances). The implementation of the whole simulation procedure is done in Matlab.

In the first simulation setup, spatial variation in energy deposition is introduced by illuminating the lens with two inclined Gaussian laser beams, each intersecting the lens at a different incident angle. An exemplary simulation result is shown in Fig.~\ref{fig:side_twobeams}, illustrating the computed temperature fields resulting from the three heat mapping strategies: (a) shape function-based projection, (b) global inverse distance weighting, and (c) element-wise inverse distance weighting. Despite of the use of a relatively coarse finite element mesh, all three methods qualitatively reproduce the expected thermal behavior. Specifically, the temperature field exhibits a gradient from the front (left) to the back (right) surface of the lens, in accordance with the attenuation of laser intensity through the material. Additionally, a localized temperature increase is observed in the center area of the beams.

A closer comparison reveals that the shape function-based method and the global inverse distance method yield similar temperature distributions, both in terms of spatial shape and peak values. This indicates that both approaches are capable of distributing the absorbed energy over the mesh nodes in a consistent and physically plausible manner. In contrast, the element-wise inverse distance method results in noticeably lower peak temperatures.

\def\tmpheight{6cm}

\begin{figure}[htbp]     
     \centering
     \begin{subfigure}[b]{0.3\textwidth}
          \centering
         \includegraphics[height=\tmpheight]{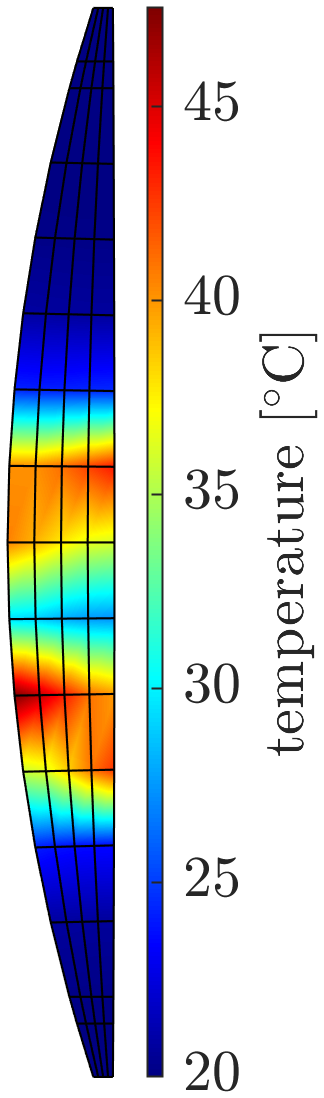}
         \caption{Mapping with shape function}
     \end{subfigure}
     \quad
     \begin{subfigure}[b]{0.3\textwidth}
          \centering
         \includegraphics[height=\tmpheight]{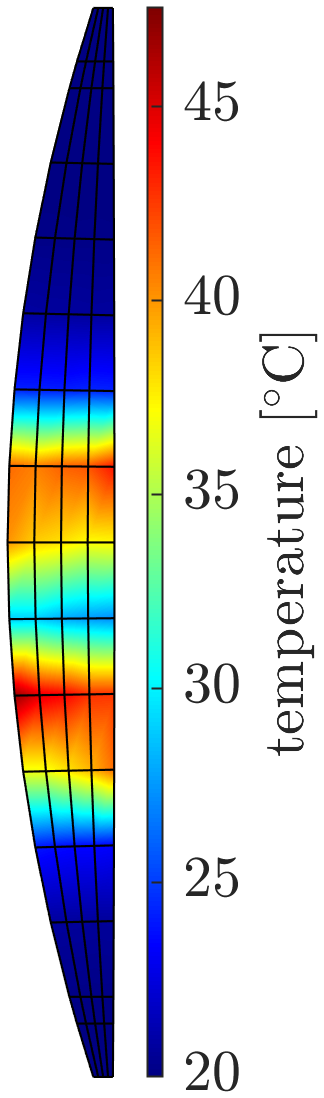}
         \caption{Mapping with global inverse distances}
     \end{subfigure}
     \quad
     \begin{subfigure}[b]{0.3\textwidth}
          \centering
         \includegraphics[height=\tmpheight]{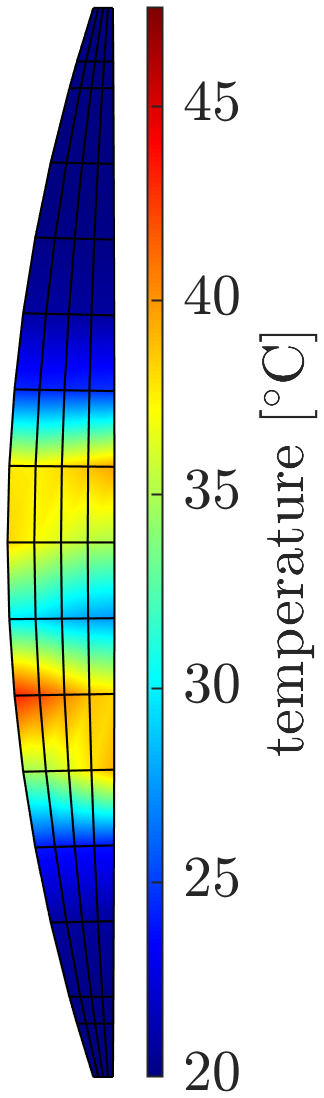}
         \caption{Mapping with element-wise inverse distances}
     \end{subfigure}
     \caption{Exemplary simulation of two transmitted Gaussian laser beams for different mapping methods while $\Nseg = 30$ point heat sources along each ray within the lens are considered.}
     \label{fig:side_twobeams}
 \end{figure}

 \begin{figure}[htbp]     
     \centering
     \begin{subfigure}[b]{0.3\textwidth}
          \centering
         \includegraphics[height=\tmpheight]{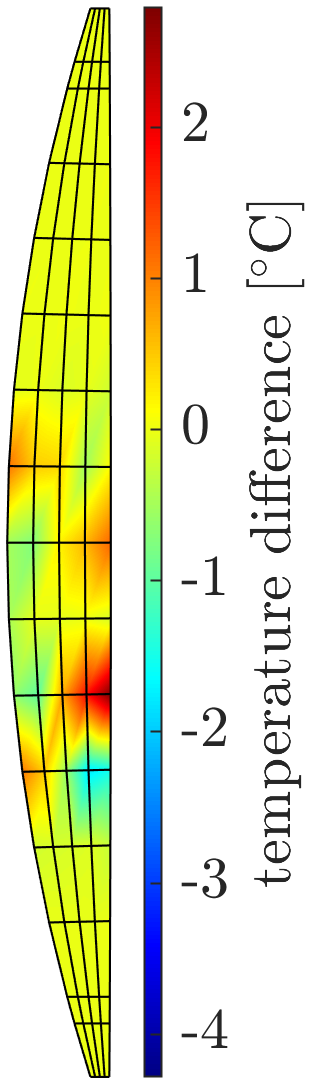}
         \caption{Mapping with shape functions vs. mapping with global inverse distances}
         \label{fig:sub1}
     \end{subfigure}
     \quad
     \begin{subfigure}[b]{0.3\textwidth}
          \centering
         \includegraphics[height=\tmpheight]{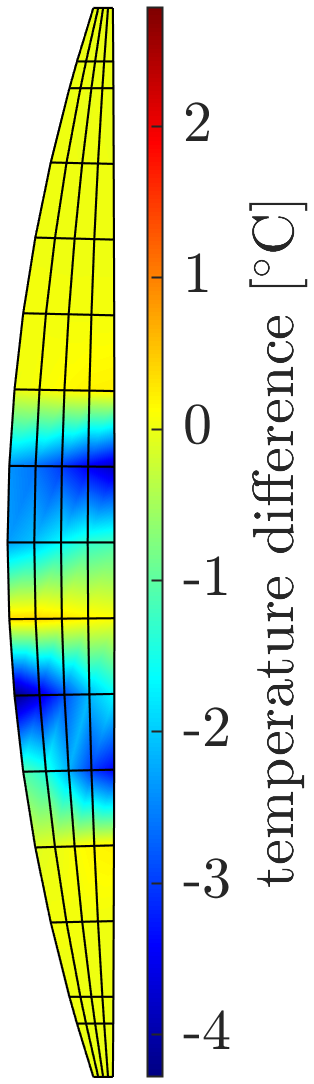}
         \caption{Mapping with shape functions vs. mapping with element-wise inverse distances}
         \label{fig:sub2}
     \end{subfigure}
     \caption{Comparing the mapping methods with inverse distances to the method with shape functions.}
     \label{fig:comp_distribtype}
 \end{figure}

A more detailed comparison of the heat mapping strategies is provided in Fig.~\ref{fig:comp_distribtype}, where the two inverse distance-based methods are directly compared to the shape function-based mapping. The plot illustrates how each method distributes the heat from the heat source over the finite element mesh and reveals differences in how localized or diffuse the resulting temperature profiles become.

It can be observed, that the mapping method using global inverse distance weighting exhibits only minor deviations from the reference solution obtained with shape functions. This suggests that global weighting provides a sufficiently smooth and consistent heat distribution across the mesh. In contrast, the element-wise inverse distance method shows more pronounced discrepancies.

\def\tmpwidth{2.4cm}
\def\tmpheight{6cm}

\begin{table}[htbp]
\centering
\begin{tabular}{>{\centering\arraybackslash}m{1.5cm}>{\centering\arraybackslash}m{\tmpwidth}>{\centering\arraybackslash}m{\tmpwidth}>{\centering\arraybackslash}m{\tmpwidth}}
     \toprule
     & $\Nseg = 3$ & $\Nseg = 10$ & $\Nseg = 20$ \\
     \cmidrule(lr){2-2}\cmidrule(lr){3-3}\cmidrule(lr){4-4}
     \textbf{shape function} & 
     \includegraphics[height=\tmpheight]{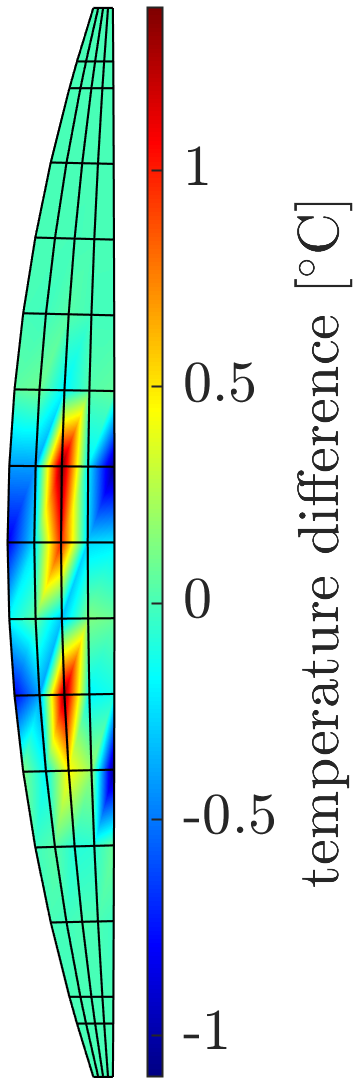} & 
     \includegraphics[height=\tmpheight]{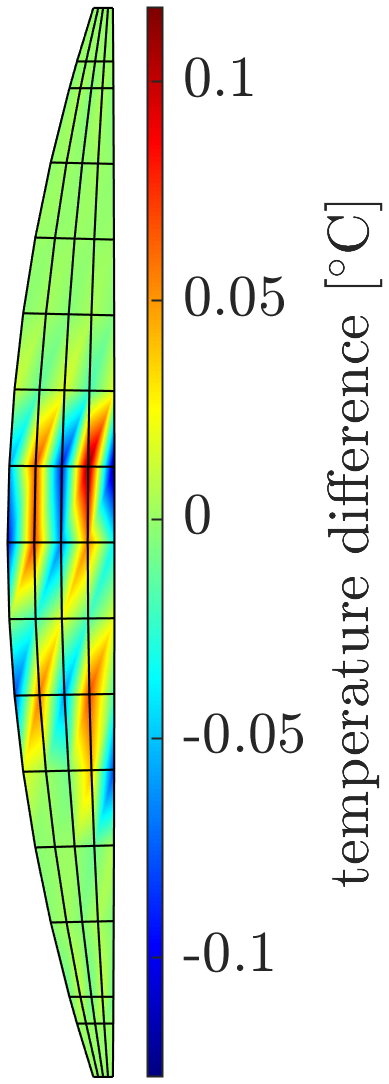} & 
     \includegraphics[height=\tmpheight]{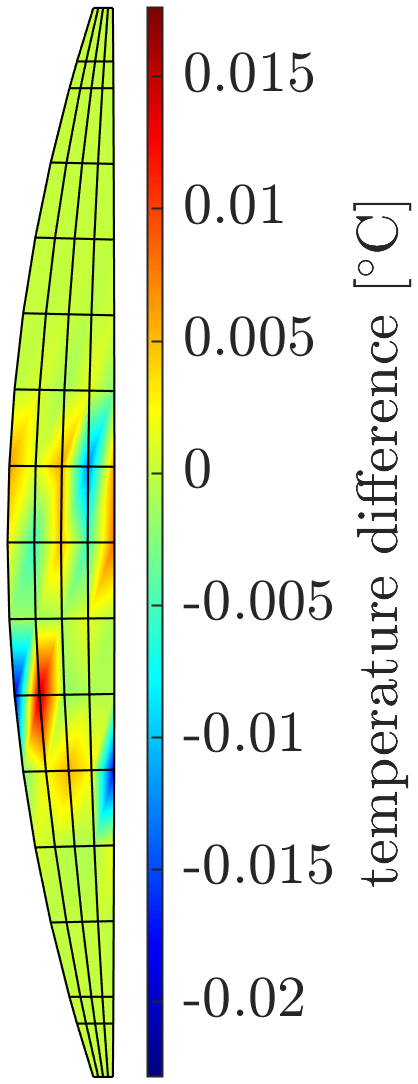} \\
     \textbf{global inverse distances} & 
     \includegraphics[height=\tmpheight]{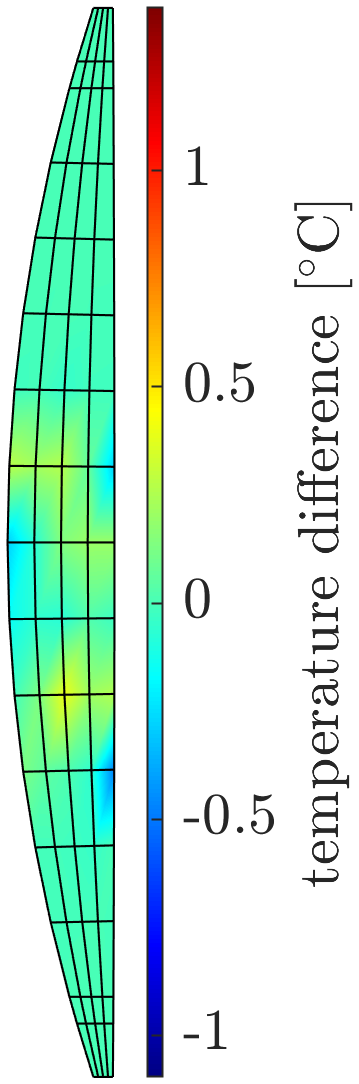} & 
     \includegraphics[height=\tmpheight]{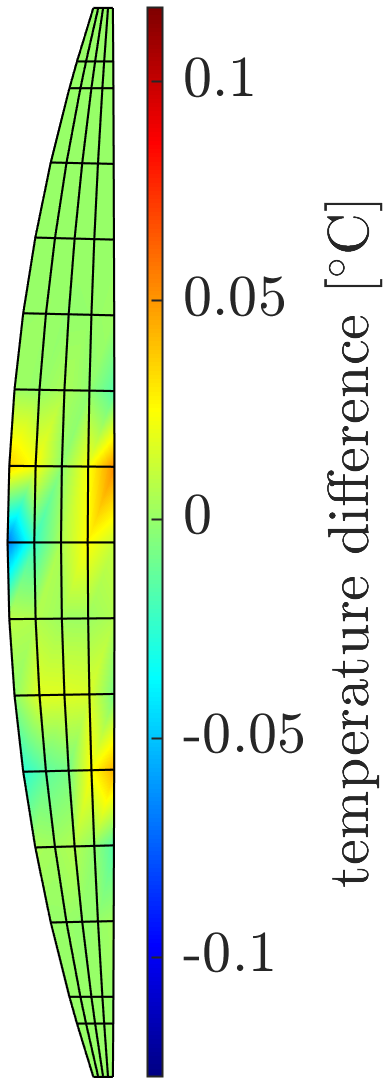} & 
     \includegraphics[height=\tmpheight]{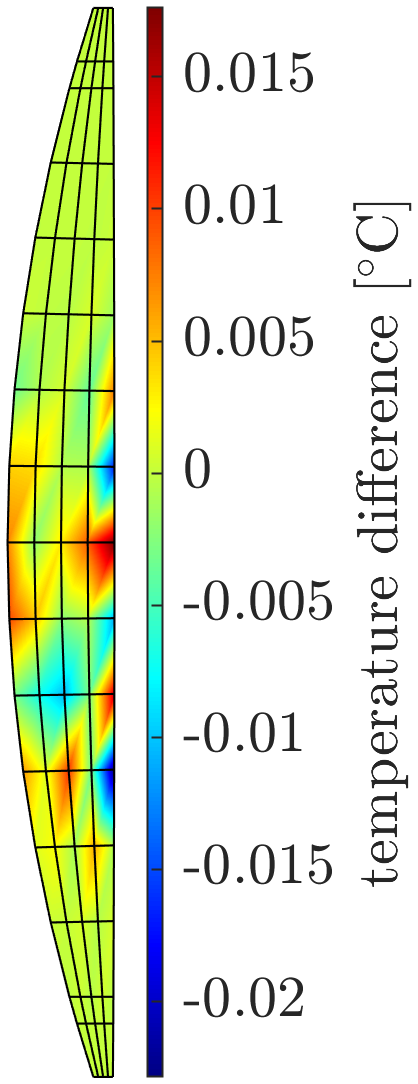} \\
     \textbf{element-wise inverse distances} & 
     \includegraphics[height=\tmpheight]{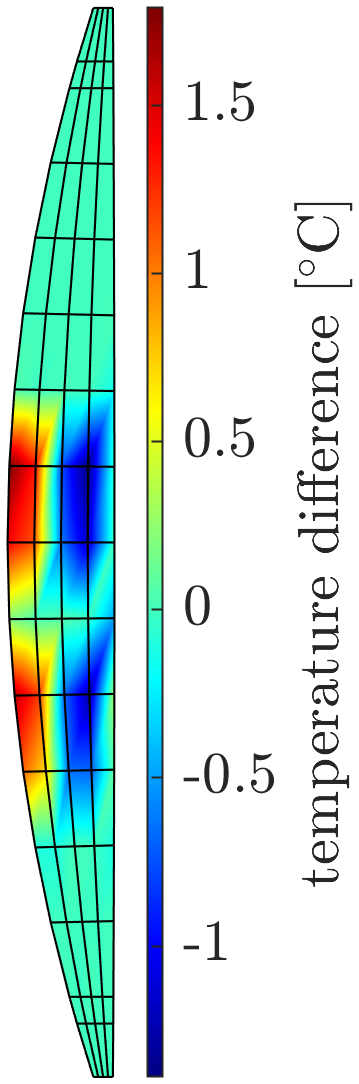} & 
     \includegraphics[height=\tmpheight]{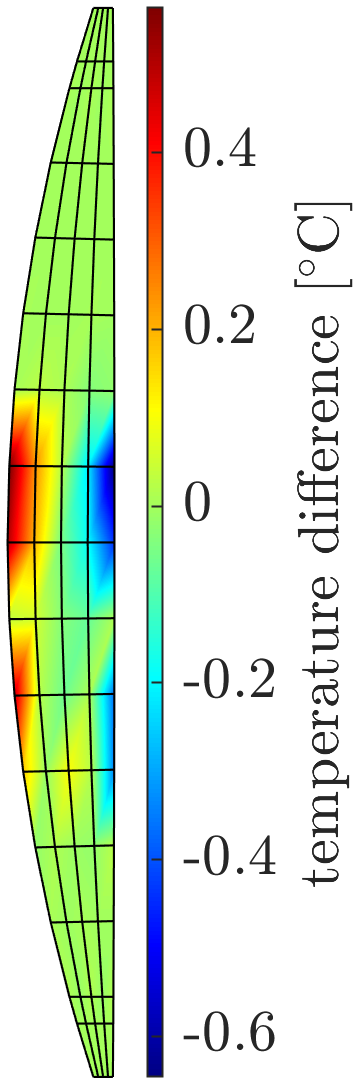} & 
     \includegraphics[height=\tmpheight]{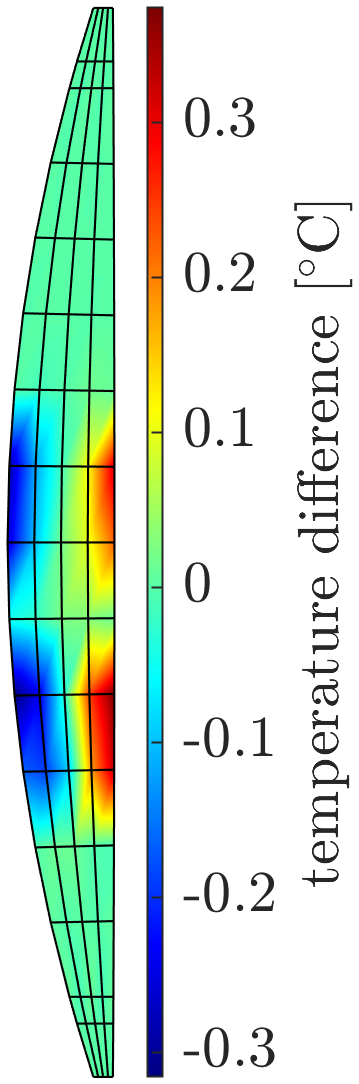} \\
     \bottomrule
\end{tabular}
\caption{Convergence study varying the number $\Nseg$ of volumetric point heat sources along a single ray for each mapping method with $\Nseg = 30$ serving as the reference solution.}
\label{table:conv_study} 
\end{table}

The influence of the number of point heat sources $\Nseg$ along each ray within the lens on the resulting temperature distribution for the three mapping procedures is analyzed in Tab.~\ref{table:conv_study}. For this purpose, $\Nseg$ is varied across several values for each mapping method, with the case of $\Nseg = 30$ serving as the reference solution.

Across all mapping strategies, it is observed that increasing the number of point heat sources improves the accuracy of the resulting temperature distribution. The temperature profiles become more consistent with the reference solution as the discretization of the heat input becomes finer. For this reason it can be concluded that it is not necessary to have a very fine discretization of the volumetric absorption, which reduces the computational effort. 

Among the three methods, the mapping approaches using shape functions and global inverse distances exhibit a faster convergence towards the reference. This indicates that these methods are capable of capturing the spatial characteristics of the heat input more effectively, even with a relatively coarse segmentation of the ray. In contrast, the element-wise inverse distance method shows slower convergence and does not reach the same level of accuracy as the other two methods. Notably, the most significant deviations occur near the entry and exit zones of the laser beams. This behavior is attributed to how the volumetric absorption is discretized and mapped within individual elements. Specifically, due to the segment-based sampling of the beam path, it can happen that a disproportionate number of point heat sources fall into the entry element compared to the exit element. This causes an imbalance in heat input, resulting in artificially elevated temperatures at the lens front and suppressed temperatures at the rear. Furthermore, in the element-wise mapping approach, heat contributions from each point source are distributed to all nodes of the enclosing element based on inverse distances. Since all distances are finite, even for nodes that are physically further away, every node within the element receives a non-negligible share of the heat input. This results in a spatially smoother heat distribution and lower local peak temperatures compared to the other methods. 

In summary, the element-wise method tends to diffuse the heat more broadly and can introduce asymmetries at the boundaries of the beam path due to localized oversampling or undersampling of segments. However, in intermediate regions of the beam path, where the number of heat sources per element tends to average out, this effect is less pronounced.

In addition to accuracy, computational performance is also a critical factor. All mapping strategies require, at a minimum, a nearest-node search for each point heat source. The global inverse distance method is computationally the least expensive, as it does not require element-specific computations beyond this initial search. On the other hand, the shape function method requires a more complex element search, which involves solving for local coordinates using the Newton-Raphson iteration. Therefore, the computation time is $317.13\%$ longer compared to the global inverse distance method. Similarly, the element-wise inverse distance method relies on identifying the enclosing element, for which a Delaunay triangulation is used. The computation time is $1237.87 \%$ longer compared to the global inverse distance method.

From the simulation results, it is found that the Newton-Raphson method for shape function mapping is computationally less expensive than the Delaunay triangulation used for element-wise mapping. However, among all tested methods, the global inverse distance mapping demonstrated the lowest overall computational effort, as it only requires identifying the nearest nodes without the need for any element search or iterative procedures. While being slightly less accurate than the shape function method in certain cases, its simplicity and speed make it particularly attractive for applications where computational resources are limited or where a fast approximation is sufficient. Therefore, depending on the required trade-off between accuracy and performance, the global inverse distance method can be a very effective alternative for efficiently modeling volumetric heat deposition in finite element simulations.

\begin{figure}[htbp]
     \centering
     \includegraphics[height=\tmpheight]{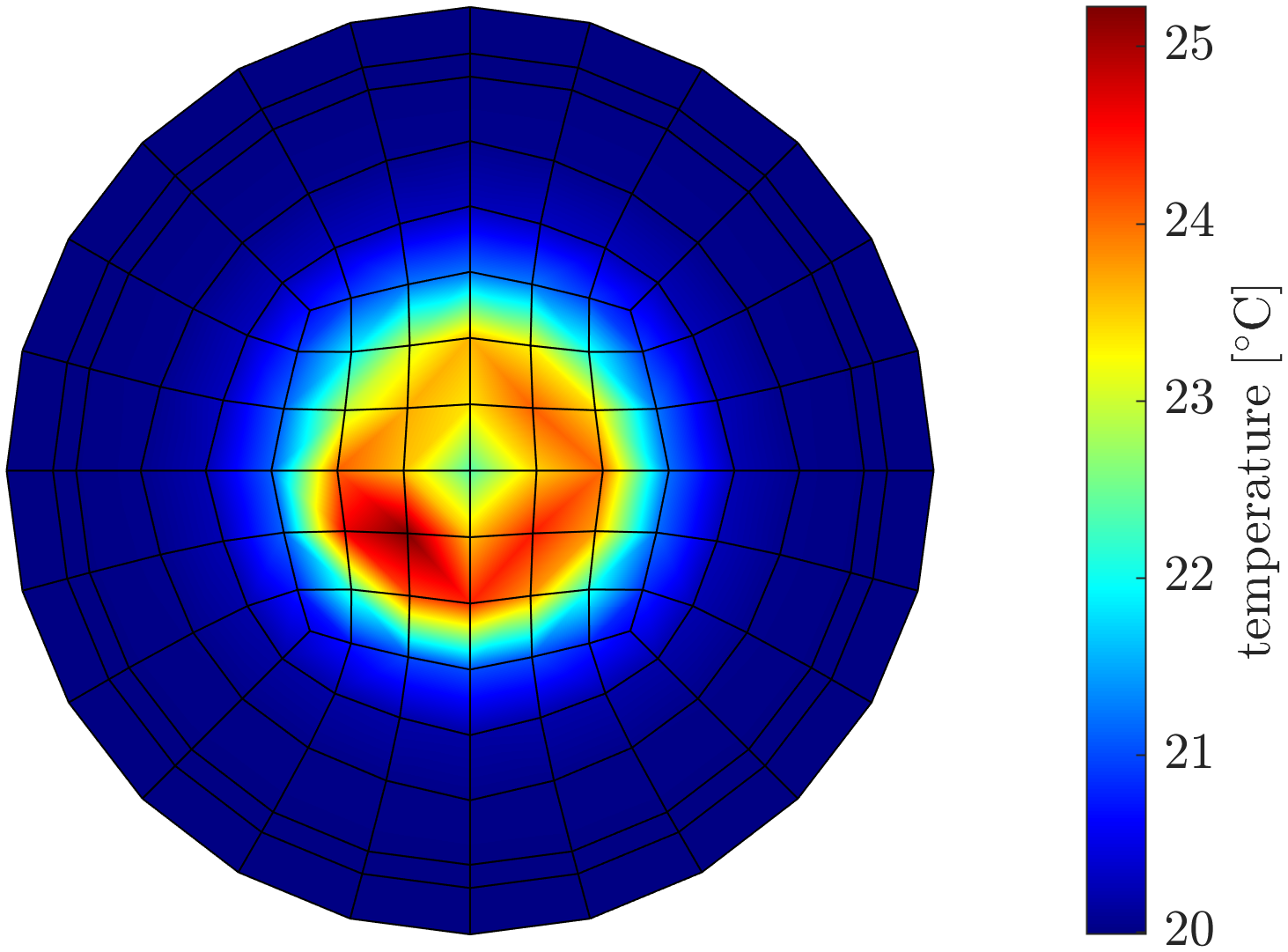}
     \caption{Transient temperature field for laser beam following a circular path.}
     \label{fig:timevarying}
\end{figure}

An exemplary simulation involving a time-varying laser beam is presented in Fig.~\ref{fig:timevarying}. In this setup, the laser beam is prescribed to move along a circular trajectory on the lens surface, introducing continuous spatial and temporal variation in energy deposition. The temperature field is computed using the previously described algorithm with the mapping method based on shape functions selected for its higher accuracy and robustness. As seen in the resulting temperature distribution, the trajectory of the laser beam is clearly reflected in the thermal footprint, showing a distinct circular heating pattern. This confirms the method's capability to resolve dynamic beam motion and transient heat input. The simulation highlights not only the spatial resolution of the approach, but also its ability to maintain stability and physical accuracy over time, making it well-suited for applications involving scanning or modulated laser sources in thermal-optical simulations.

\section{Conclusion}
\label{sec:conclusion} 

This paper presents an algorithm for modeling both surface and volumetric laser absorption in optical elements, with an emphasis on methods for mapping point heat sources along each ray to finite element nodes. The proposed framework offers high flexibility in handling spatial and temporal variations in beam parameters and element positioning, making it well-suited for simulating realistic optical systems under complex illumination conditions.

Among the mapping methods evaluated, the global inverse distance approach demonstrates the lowest computational cost while producing temperature distributions that closely resemble those obtained from the more accurate shape function-based method. This makes it a promising alternative for time-sensitive simulations where computational efficiency is a higher priority than peak accuracy.

%
%
%


\end{document}